\documentclass[aps,prd,reprint,preprintnumbers,floatfix,superscriptaddress]{revtex4-2}

\usepackage{graphicx} 
\usepackage{amsmath}
\usepackage{units}

\usepackage{etoolbox}
\makeatletter
\patchcmd{\frontmatter@RRAP@format}{(}{}{}{}
\patchcmd{\frontmatter@RRAP@format}{)}{}{}{}
\renewcommand\Dated@name{}
\makeatother

\begin{document}
\title{The ICEBERG Test Stand for DUNE
Cold Electronics Development}
\collaboration{Contribution to the 25th International Workshop on Neutrinos from Accelerators}

\author{Alejandro Yankelevich}
\affiliation{University of California at Irvine, Irvine, California 92697, USA}
\author{for the DUNE collaboration}
\noaffiliation

\preprint{FERMILAB-CONF-24-0874-LBNF}

\date{Presented September 19, 2024}

\begin{abstract}
ICEBERG is a liquid argon time projection chamber at Fermilab for the purpose of testing detector components and software for the Deep Underground Neutrino Experiment (DUNE). The detector features a \unit[1.15]{m} x \unit[1]{m} anode plane following the specifications of the DUNE horizontal drift far detector and a newly installed X-ARAPUCA photodetector. The status of ICEBERG is reported along with analysis of noise, pulser, and cosmic ray data from the ninth run beginning May 2024 with the goal of advising the DUNE collaboration on the optimal wire readout electronics configuration. In addition, development of an absolute energy scale calibration method is currently underway using known sources such as cosmic ray muon Michel electrons at the $\sim$\unit[10]{MeV} scale and $^{39}$Ar decay electrons at the $\sim$\unit[100]{keV} scale. Research into AI-based identification of such events at the data acquisition level is introduced.\par
\end{abstract}
\maketitle

\section{Introduction}
DUNE \cite{dune} is a long-baseline neutrino oscillation experiment that will use Fermilab’s future PIP-II beam. Four \unit[17]{kt} liquid argon time projection chamber (LArTPC) modules are planned for the far detector at SURF in South Dakota including two modules that will be ready during the first phase of the experiment: a horizontal drift module and a vertical drift module. The horizontal drift module will use wire readout technology in which ionized drift electrons are detected through three wire planes at the anode plane assemblies (APAs).\par
ProtoDUNE \cite{protodune} is a prototype of the horizontal drift and vertical drift modules built in the neutrino platform at CERN in order to test the LArTPC technology to be used in DUNE before the full far detectors are built. These smaller \unit[0.77]{kt} detectors take data from cosmic rays and a dedicated beamline from CERN's Super Proton Synchrotron. ProtoDUNE previously ran from 2018-2019, and the horizontal drift module started its second run in May 2024, which included eight weeks of beam runs starting mid-June. The vertical drift module will start running in early 2025.\par

\subsection{ICEBERG}
\begin{figure}[t]
    \begin{tabular}{cc}
        \includegraphics[width=1.7in]{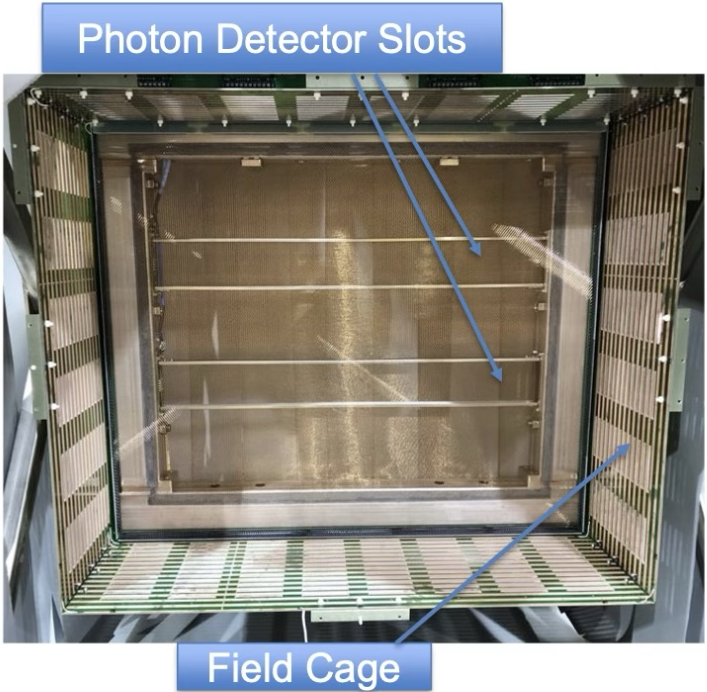}
        \includegraphics[width=1.6in]{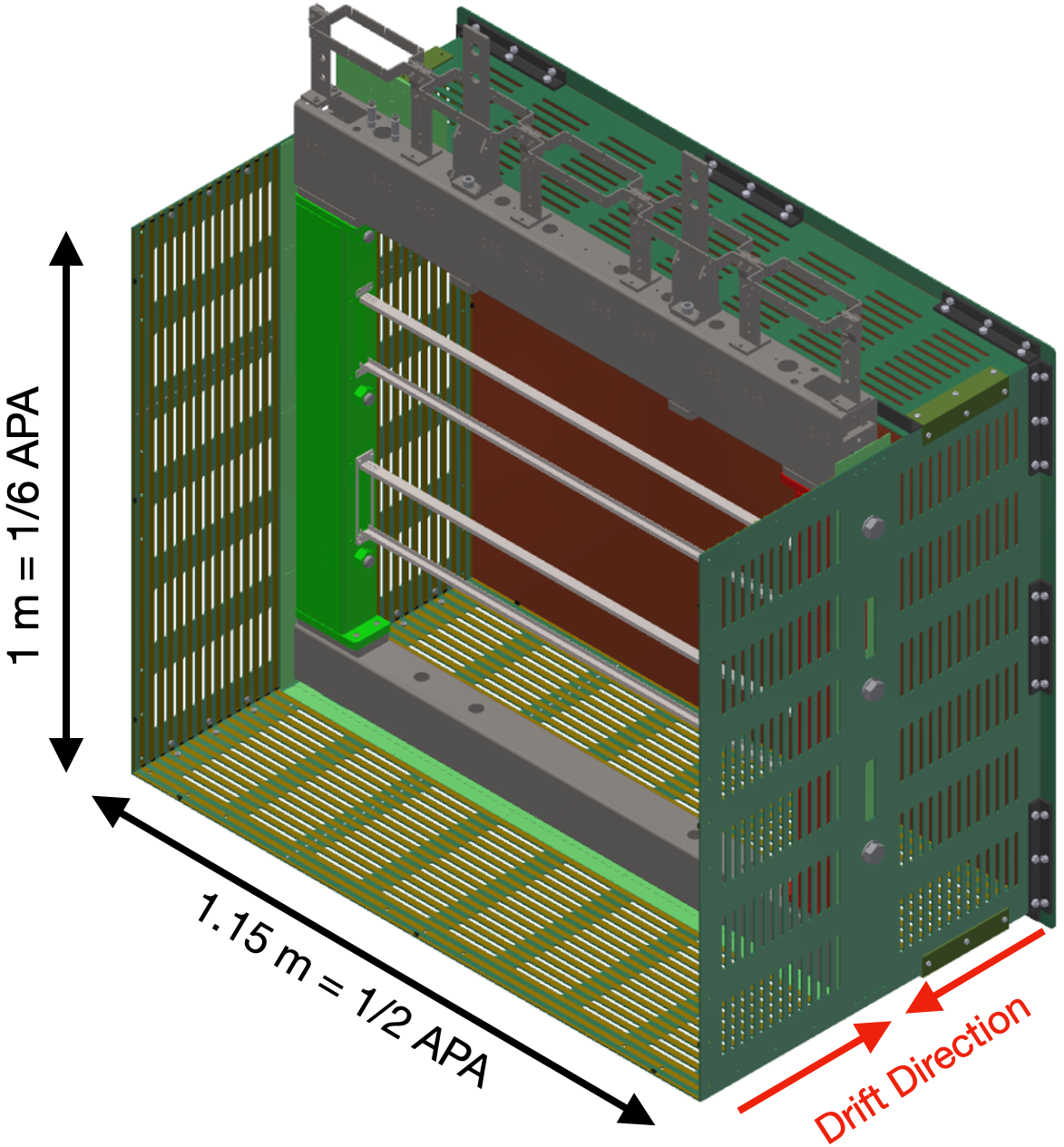} 
    \end{tabular}
    \caption{
        Picture and schematic of ICEBERG TPC. One APA is located in the middle of the active region with cathode planes on opposite ends. There are slots for two ARAPUCA photodetectors on the APA.
    }
    \label{fig:iceberg}
\end{figure}

\begin{figure}[t]
    \includegraphics[width=2in]{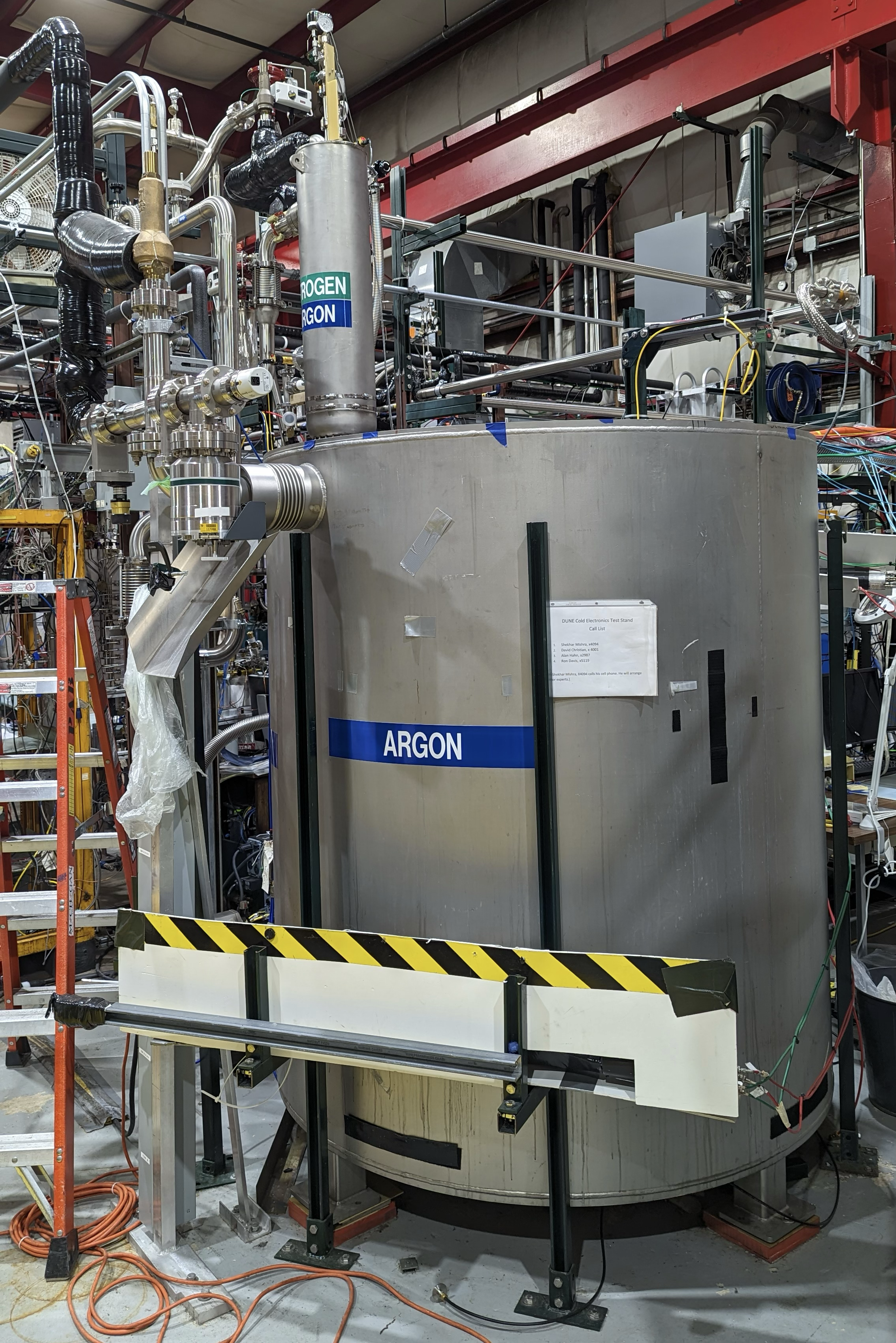} 
    \caption{
        The ICEBERG cryostat. One of two scintillator bars for cosmic triggering is visible at the front bottom of the cryostat.
    }
    \label{fig:cryostat}
\end{figure}

The Integrated Cryostat and Electronics Built for Experimental Research Goals (ICEBERG) at Fermilab is a smaller prototype of the DUNE horizontal drift far detector module and is intended for tests of the cold TPC readout electronics and photodetection system. It consists of two \unit[1.15]{m} x \unit[1.00]{m} x \unit[0.30]{m} drift volumes (Fig.~\ref{fig:iceberg}). ICEBERG primarily takes cosmic ray data, and there are two scintillator bars on opposite corners of the cryostat in order to trigger on cosmics traveling parallel to a diagonal wire plane (Fig.~\ref{fig:cryostat}).\par
A full-size APA used in DUNE and ProtoDUNE is \unit[2.3]{m} x \unit[6.2]{m} and consists of 2560 read out wire channels distributed among two diagonal planes and one vertical plane. The DUNE horizontal drift detector will contain 150 APAs, and ProtoDUNE uses 4 APAs. In comparison, ICEBERG has a single \unit[1.15]{m} x \unit[1.00]{m} APA with 1280 channels. The motivation for this APA geometry with 1/2 the width but 1/6 the height of a full APA is to maximize the number of channels in this small detector size while maintaining identical wire spacing: since one wire plane consists of vertical wires, reducing the width proportionally reduces the number of channels.\par
\begin{figure}[t]
    \includegraphics[width=2.3in]{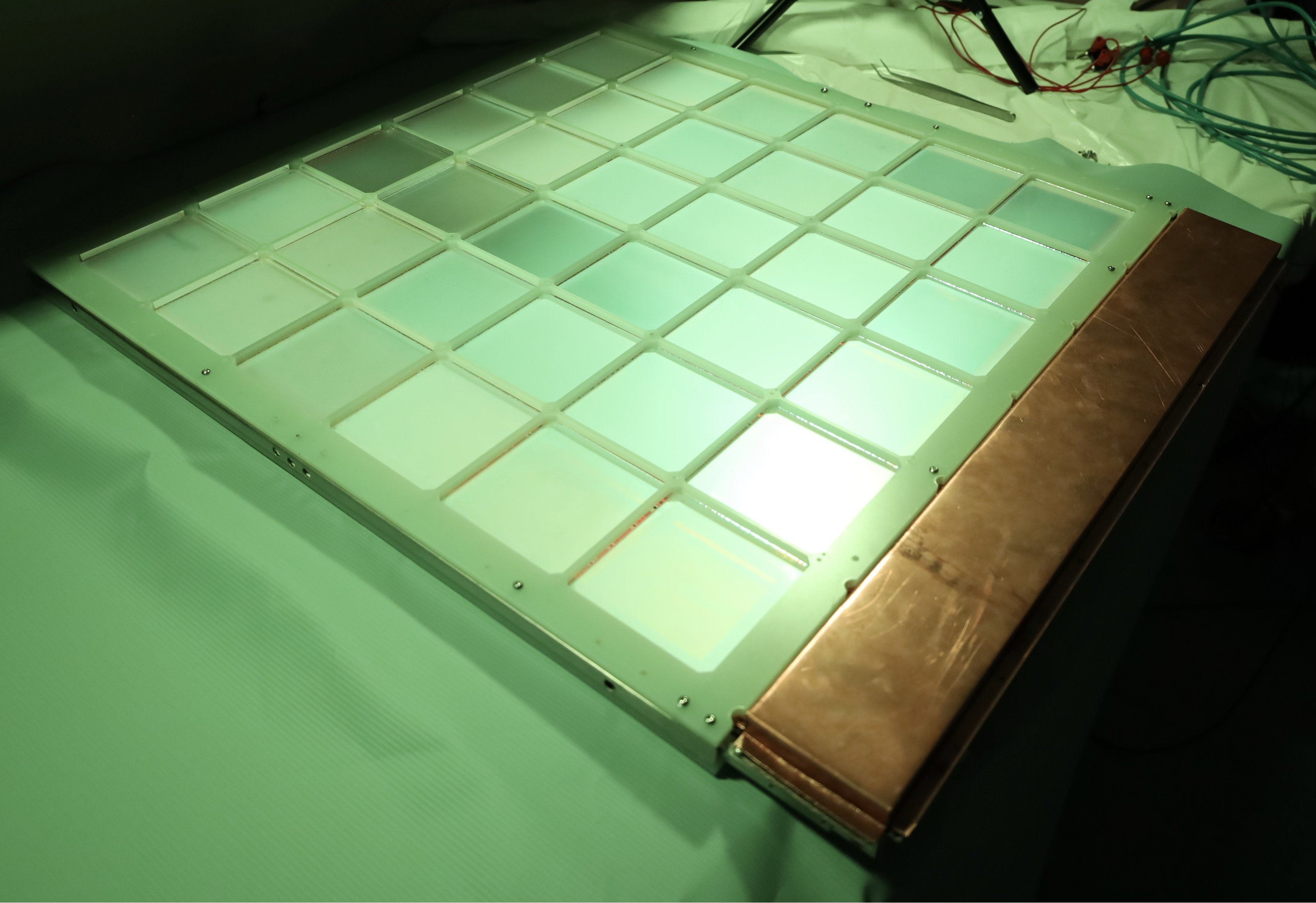} 
    \caption{
        An X-ARAPUCA photodetector for the DUNE vertical drift module.
    }
    \label{fig:x_arapuca}
\end{figure}

During previous runs of ICEBERG including the most recent of which ending in 2019, two horizontal drift ARAPUCA photodetectors \cite{dune_iv}  were installed in the ICEBERG APA (Fig.~\ref{fig:iceberg}). In the current run, however, an upgraded vertical drift X-ARAPUCA \cite{x_arapuca} (Fig.~\ref{fig:x_arapuca}) was installed beneath the bottom TPC field cage. In addition to testing this particular photodetector model in a TPC environment for the first time, this allows for testing the power and signal over fiber capabilities necessary for the photodetectors placed on the DUNE vertical drift cathodes, which will be at \unit{-300}{kV} \cite{pof}.\par


\subsection{Goals}
The main motivation of the current ProtoDUNE run is to validate DUNE's physics goals with a single consistent run setting rather than to test detector elements under various run configurations. However, ICEBERG can be used to conduct several tests of detector elements with shorter commissioning time between runs due to its small size and is available to take data under various run configurations in order to determine the optimal settings for the cold electronics. The current goals of ICEBERG are therefore to (1)~test the latest versions of the DUNE cold electronics, which feature minor improvements compared to those installed in ProtoDUNE, (2)~test the vertical drift X-ARAPUCA photodetector, (3)~test a new detector safety system that turns off the high voltage to the wire planes if the cold electronics are powered off and turns off the warm electronics outside the detector if the temperature gets too hot or the fans are not powered, (4)~advise DUNE in addition to ProtoDUNE before it had started its beam runs on the optimal choice of cold electronics readout settings, and (5)~to develop an AI-based method for online identification of possible calibration sources such as $^{39}$Ar decays and Michel electrons to provide an absolute calibration of the electronics.\par

\section{Commissioning}
\begin{figure}[t]
    \includegraphics[width=3.2in]{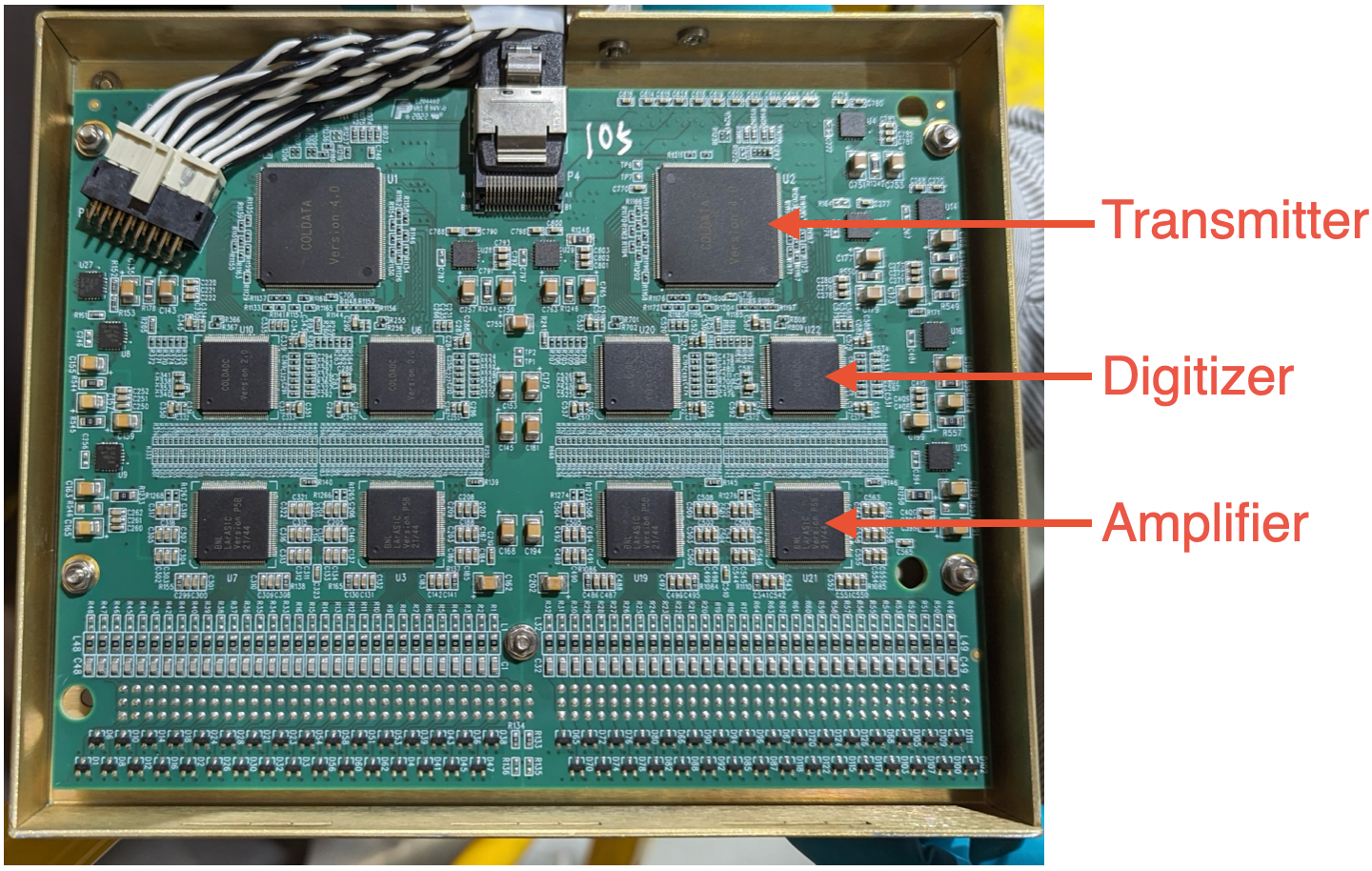} 
    \caption{
        A front-end mother board. Analog wire signals input from the bottom are processed through an amplifier chip, digitizer chip, and transmitter chip that communicates with warm electronics outside of the cryostat.
    }
    \label{fig:femb}
\end{figure}

The commissioning of ICEBERG for its current run began in November 2023 upon the receipt of the latest versions of the front end mother boards (FEMBs), which house the cold electronics \cite{dune_iv}. These FEMBs consist of amplifier \cite{larasic}, digitizer \cite{coldadc}, and transmitter \cite{coldata} chips and are responsible for reading out the TPC wire signals (Fig.~\ref{fig:femb}). While the designs are being continuously updated, the model in ICEBERG is intended to be the final version to be used in DUNE and features minor updates compared to the one currently installed in ProtoDUNE including improvements to the circuit layout, cable clamps, and screw holes.\par

Before installation on the ICEBERG APA, these FEMBs were first cabled to the external warm electronics and tested in liquid nitrogen inside of a cryogenic test stand \cite{dune_iv}. During these tests, noise data was taken through ICEBERG’s DAQ to ensure all FEMBs were functional. Following the FEMB tests, the ICEBERG commissioning process continued, which included mounting the FEMBs onto the top of the APA and attaching the vertical drift photodetector beneath the TPC. The TPC was lifted into the ICEBERG cryostat in April 2024, and the subsequent filling with liquid argon completed in May.\par

\section{Cold Electronics Tests}
\begin{figure}[t]
    \includegraphics[width=2.7in]{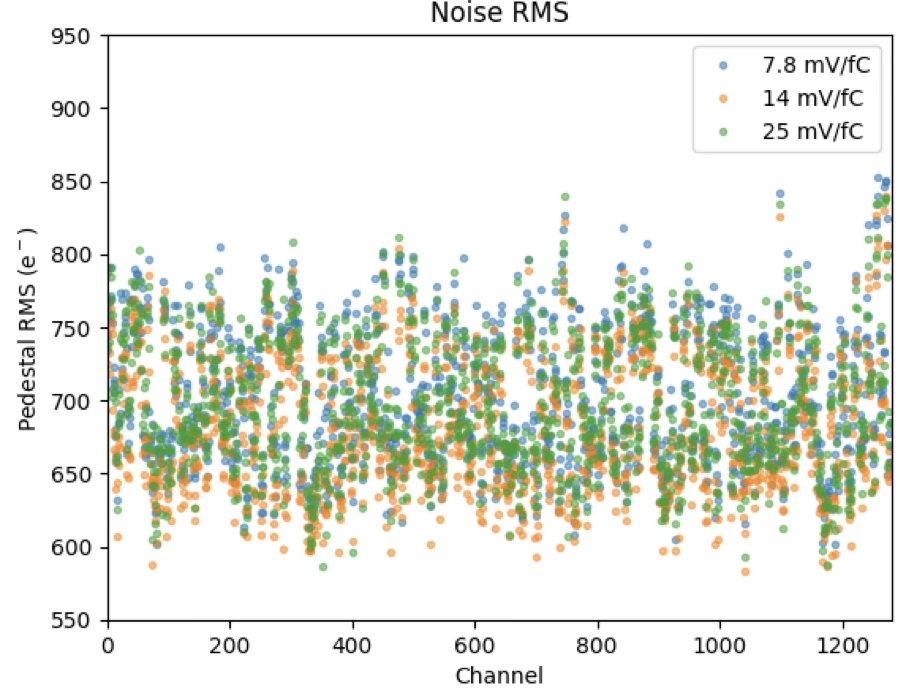} 
    \includegraphics[width=2.7in]{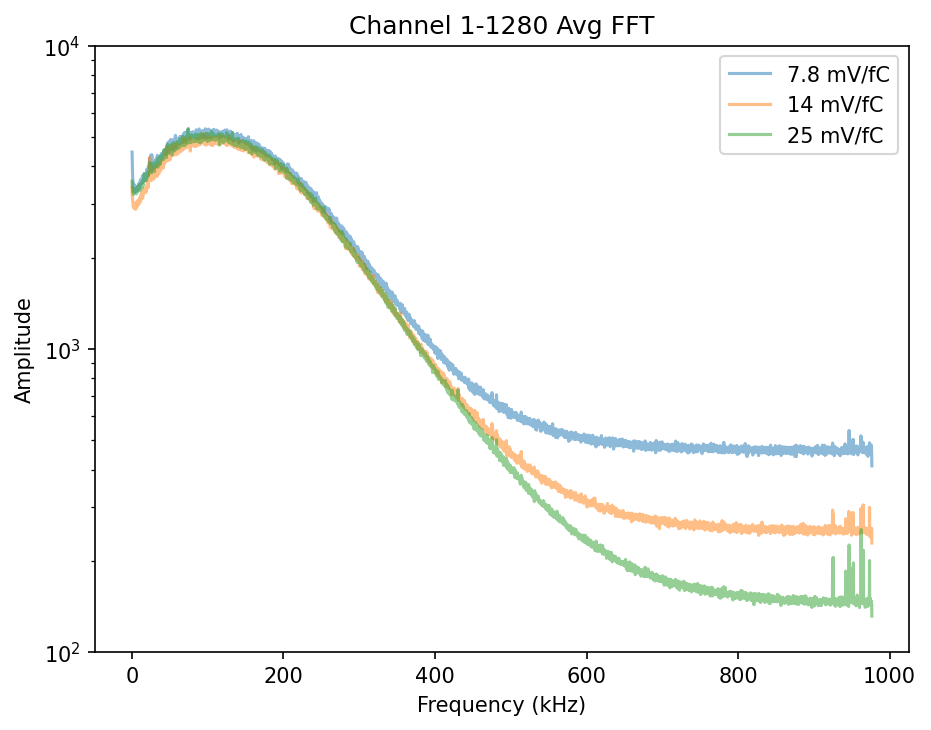} 
    \caption{
        Channel waveform RMS converted to equivalent drift electrons (top) and average FFT over all 1280 channels (bottom) under \unit[7.8]{mV/fC}, \unit[14]{mV/fC}, and \unit[25]{mV/fC} gain settings. FFT amplitude units are arbitrary. The ICEBERG TPC here is in the cryostat and at room temperature. Similar RMS levels are observed across the three settings, and there are few features observed in the three FFT plots.
    }
    \label{fig:gain}
\end{figure}

A major goal of this ICEBERG run was to find the ideal settings for the cold electronics, meaning those that give the best signal to noise ratio. The settings that are particularly of interest include the gain of the amplifiers, the shaping time of the amplifiers, the baseline of the digitizers, and the voltage applied to the voltage regulators that power each of the three types of chips. As an example, noise tests in which the gain setting was varied are shown in Fig.~\ref{fig:gain}.

This test shows that noise levels are comparable between the three possible gain settings as can be seen through the comparable waveform RMS and featureless FFT among these gain settings. Since a lower gain setting provides a greater dynamic range and avoids saturating the amplifier, the lower \unit[7.8]{mV/fC} setting was chosen as optimal. This setting was therefore also adopted as nominal for the current ProtoDUNE run, changing from the previous nominal choice of \unit[14]{mV/fC}.\par

\begin{figure}[t]
    \includegraphics[width=2.7in]{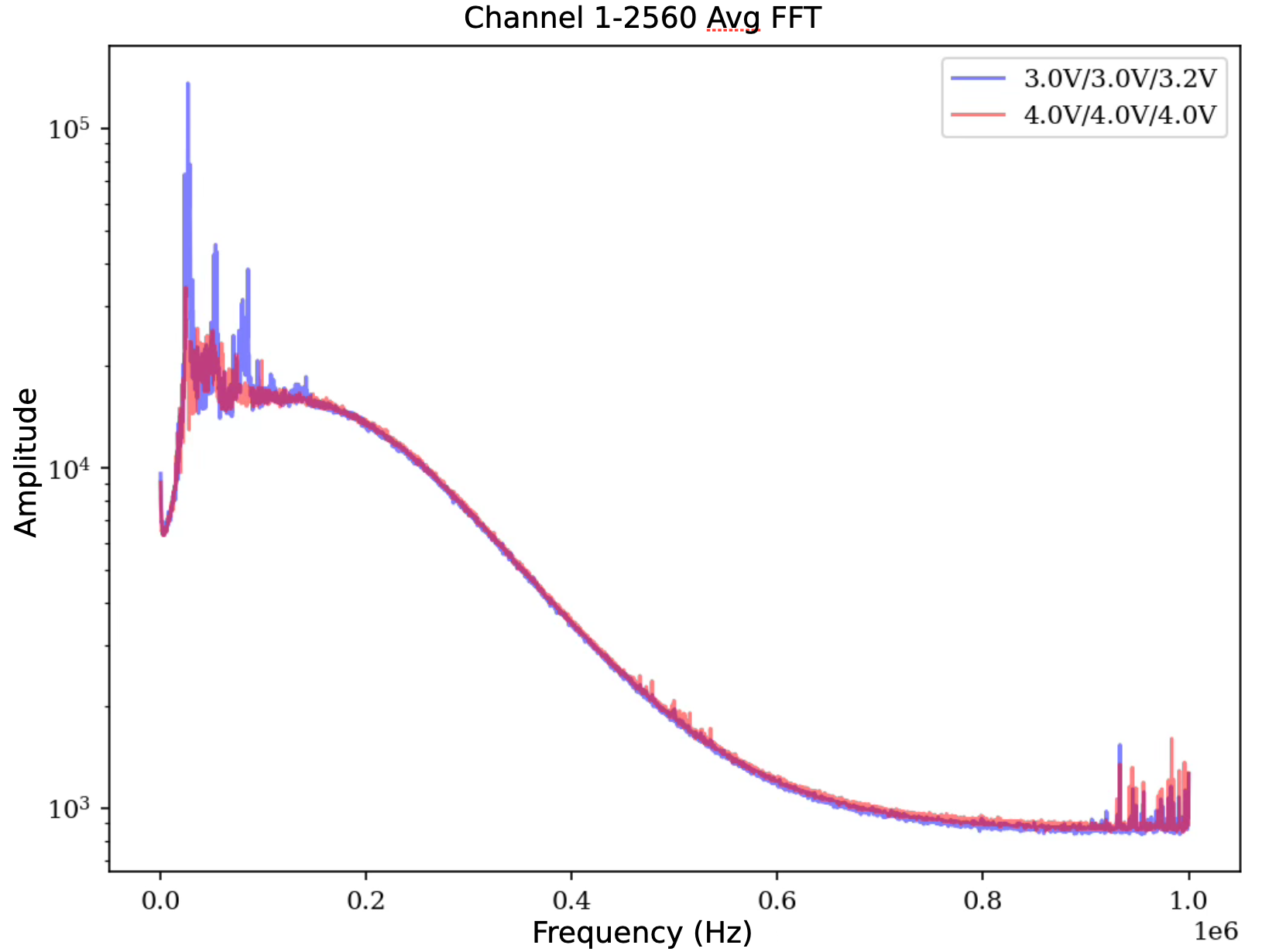} 
    \includegraphics[width=2.7in]{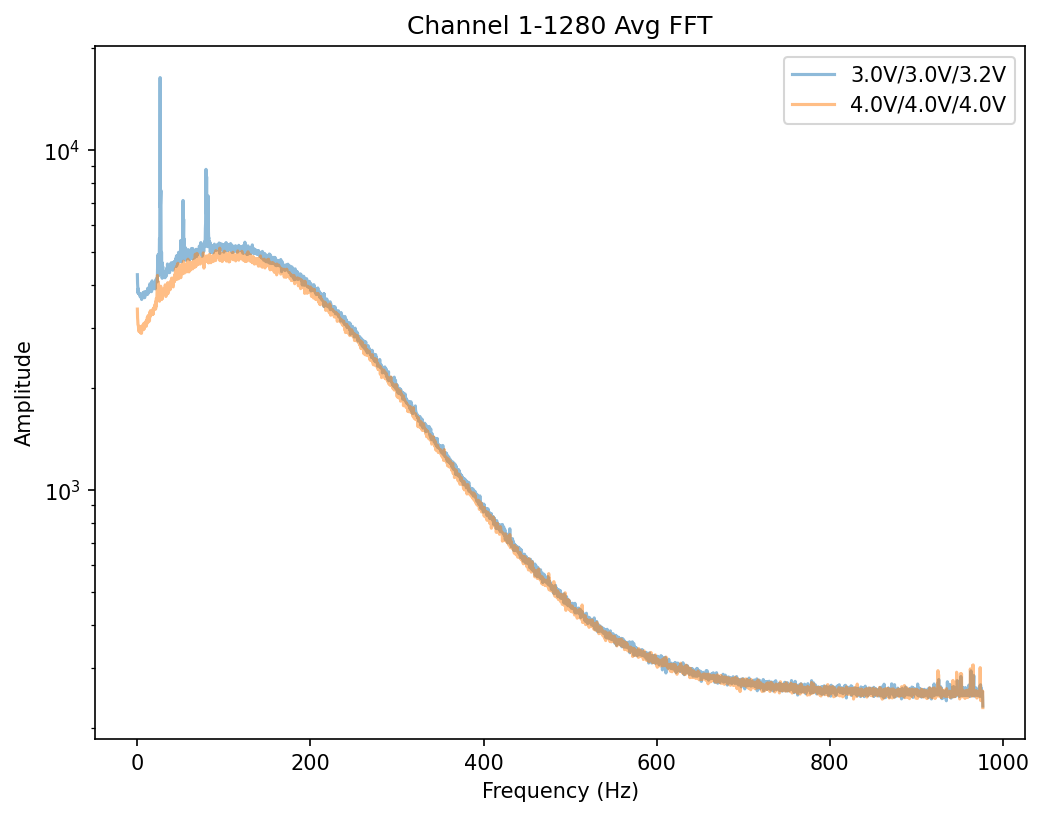}
    \caption{
        Average FFT over all channels under \unit[3.0]{V}/\unit[3.0]{V}/\unit[3.2]{V} and \unit[4.0]{V}/\unit[4.0]{V}/\unit[4.0]{V} power settings to the amplifier/transmitter/digitizer voltage regulators at ICEBERG inside its cryostat (bottom) and one ProtoDUNE APA inside a coldbox (top), both at room temperature. Peaks at \unit[25]{kHz}, \unit[50]{kHz}, and \unit[75]{kHz} are observed for both detectors' APAs, but they are reduced at the \unit[4.0]{V} settings and nearly absent in ICEBERG's case. The orange plot in the bottom figure is from the same test as the orange plot in Fig.~5 (bottom).
    }
    \label{fig:power}
\end{figure}

Another highlight of the cold electronics configuration tests is regarding the power settings. The recommended voltages to supply to the voltage regulators for the transmitter, amplifier, and digitizer chips were \unit[3.0]{V}, \unit[3.0]{V}, and \unit[3.2]{V}, respectively. However, during tests of ProtoDUNE APAs, a large \unit[25]{kHz} coherent noise pickup had been observed along with its second and third harmonics (Fig.~\ref{fig:power}). A possible explanation for this noise is pickup between the power and data cables to the FEMBs. It was later seen that changing the power settings to \unit[4.0]{V} for all three types of chips largely reduced this \unit[25]{kHz} pickup. This noise is visible in ICEBERG as well despite the fact that ICEBERG has much shorter cables to the FEMBs. In ICEBERG's case, this noise is essentially absent under the alternative \unit[4.0]{V} settings, confirming that this configuration should be adopted as nominal.\par

\section{$\nu$-{O\lowercase{n}E\lowercase{dge}}}
In addition to the goals of testing detector elements, the ICEBERG DAQ is being used to develop a method for the absolute calibration of the electronics. Potential well-known calibration sources include $^{39}$Ar decays and Michel electrons from cosmic ray muons. $^{39}$Ar decays for calibration in the order of $\sim$\unit[100]{keV} are a good calibration source due to their abundance in liquid argon \cite{ar39}. For higher energies, a method to calibrate the electron energy scale with Michel electrons was demonstrated during the first run of ProtoDUNE by implementing a correction to electron energy estimation based on the theoretical Michel electron spectrum \cite{michel}.\par
An online AI-based method for such calibration event identification is being developed to run on the DAQ server GPU and would rely on ``trigger primitives" produced early in the DAQ pipeline containing basic information such as the time, width, and peak of each pulse for each hit \cite{dune_iv}. Such a method can also be extended to online particle classification, which, together with online calibration for energy estimation, would be useful for triggering on low energy events ($\unit[<10]{MeV}$) that would otherwise have lower triggering efficiency.\par

\section{Summary}
ICEBERG has started a new run currently ongoing since May 2024 with the aim of testing the latest versions of the cold electronics, the current version of the DAQ software, and a new vertical drift photodetector. ICEBERG primarily takes cosmic ray data in addition to data for specific studies such as noise tests under various run settings as well as runs with external pulses for calibration. These noise tests helped to advise DUNE and ProtoDUNE on the ideal run settings before ProtoDUNE begain receiving beam. Finally, research is currently ongoing using the ICEBERG DAQ to develop online, AI-based methods for identification of calibration events to provide an absolute electronics calibration and for particle classification. A new run with this AI DAQ integration is currently planned to start in Winter 2025.\par

\begin{acknowledgments}
This document was prepared by the DUNE collaboration using the resources of the Fermi National Accelerator Laboratory (Fermilab), a U.S. Department of Energy, Office of Science, Office of High Energy Physics HEP User Facility. Fermilab is managed by Fermi Research Alliance, LLC (FRA), acting under Contract No. DE-AC02-07CH11359. The U.S. Government retains and the publisher, by accepting the article for publication, acknowledges that the U.S. Government retains a non-exclusive, paid-up, irrevocable, world-wide license to publish or reproduce the published form of this manuscript, or allow others to do so, for U.S. Government purposes.\par
This material is based upon work supported by the U.S. Department of Energy, Office of Science, Office of Workforce Development for Teachers and Scientists, Office of Science Graduate Student Research (SCGSR) program. The SCGSR program is administered by the Oak Ridge Institute for Science and Education (ORISE) for the DOE. ORISE is managed by ORAU under contract number DE-SC0014664. All opinions expressed in this paper are the author’s and do not necessarily reflect the policies and views of DOE, ORAU, or ORISE.
\end{acknowledgments}

\bibliographystyle{unsrt}
\bibliography{main}

\end{document}